\documentclass[useAMS]{gMOS2e}

\usepackage{natbib}
\usepackage{graphicx}

\title{Ergodicity of the Nos\'e-Hoover method}

\author{
Hiroshi Watanabe$^\ast$\thanks{$^\ast$Corresponding
author. Email: hwatanabe@is.nagoya-u.ac.jp}\\\vspace{6pt}
Department of Complex Systems Science, Nagoya University, Nagoya 464-8601, Japan\\ \vspace{6pt}
Hiroto Kobayashi\\\vspace{6pt}
Department of Natural Science and Mathematics, Chubu University, Kasugai 487-8501, Japan
}

\received{v3.2 released March 2006}

\begin{document}
\label{firstpage} \doi{10.1080/0892702YYxxxxxxxx}

%\issn{1029-0435}  \issnp{0892-7022} %\jvol{00} \jnum{00} \jyear{2005} \jmonth{20 January}

\maketitle

\begin{abstract}
Ergodicity of the systems with Nos\'e-Hoover thermostat are studied.
The dynamics of the heatbath variables are investigated and
they can be periodic when the system has quick oscillation.
The periodic behaviour of them causes the system to lose its ergodicity.
The kinetic-moments method is also studied, and
the heatbath variables in this method are found to be chaotic.
The chaotic behaviour makes the whole system ergodic.
\end{abstract}

\begin{keywords}
Nos\'e-Hoover thermostat, heatbath
\end{keywords}\bigskip

\section{Introduction}

Molecular dynamics (MD) simulations have become more powerful tools with 
the development in performance of computers.
The standard MD simulations have been performed with constant energy,
and therefore, these simulations have taken samples in the microcanonical ensemble.
It is required to perform constant-temperature MD \cite{NoseReview},
because the most of physical quantities are observed in constant-temperature environment.
Nos\'e proposed the extended system method which allows to simulate
constant-temperature MD conserving time-reversible property~\cite{Nose}.
Hoover reformulated the Nos\'e's method to a more practical one,
which is now called the Nos\'e-Hoover method~\cite{NoseHoover}.

Recently, the Nos\'e-Hoover thermostat has been applied not only for
the equilibrium system but also for non-equilibrium phenomena.
Ogushi {\it et al.} studied the heat conduction of the three-dimensional
Lennard-Jones particle system using molecular dynamics simulation~\cite{Ogushi}.
They put two thermostat with different temperatures on the both ends of the system
to achieve the spontaneous phase separation.

%---
% Ergodicity of the NH system
%---

The Nos\'e-Hoover method achieves Gibbs' canonical ensemble when the
system is ergodic. However, some numerical researches report that 
the system with a small degree of freedom often loses its ergodicity,
and consequently, the system cannot be canonical in Gibbs' sense.
In the system studied by Ogushi {\it et al.}, for instance, 
the density at the higher-temperature side becomes dilute,
and the number of particles controlled by the thermostat can be 
very small. Therefore, the system can lose the ergodicity.

In order to improve the ergodicity of a system with thermostat, 
many methods have been proposed~\cite{NoseHooverChain, KineticMoments}.
However, there are less studies investigating why and how the system
loses its ergodicity.
In the present article, we focus the ergodicity of some systems in which 
temperature controlled by the  Nos\'e-Hoover method.

\section{Nos\'e-Hoover Method}

Consider an isolated system described by a Hamiltonian ${\cal H}({\bf p},{\bf q})$
with momenta {\bf p} and coordinates ${\bf q}$.
The equations of motion are
\begin{eqnarray}
\dot{q_i} &=& \frac{\partial \cal H}{\partial p_i},\\
\dot{p_i} &=& - \frac{\partial \cal H}{\partial q_i},
\end{eqnarray}
where $\dot{A}$ denotes the differentiation of $A$ with respect to time.
The energy of the system is conserved.

Following Nos\'e and Hoover, 
we can achieve the canonical ensemble by modifying the equations of motion as follows:
\begin{eqnarray}
\dot{q_i} &=& \frac{\partial \cal H}{\partial p_i}, \label{eq_nh_motion_q} \\
\dot{p_i} &=& - \frac{\partial \cal H}{\partial q_i} - \zeta p_i,\label{eq_nh_motion_p} \\
\dot{\zeta} &=& \left(\frac{K}{K_0} - 1 \right) \frac{1}{\tau^2}, \label{eq_nh_motion_zeta} \\
\dot{\eta} &=& \zeta,
\end{eqnarray}
where $\zeta$ and $\eta$ are additional variables expressing the Nos\'e-Hoover thermostat,
and $K$, $K_0$, and $\tau$ are the kinetic energy, the aimed temperature,
and the relaxation time of the thermostat, respectively.

While the system with the Nos\'e-Hoover thermostat does not have 
the canonical form, a quantity 
\begin{equation}
{\cal H'} = {\cal H} + K_0(\tau \zeta^2+2\eta)
\end{equation}
is conserved with the original Hamiltonian ${\cal H}$.
The energy of the system without the thermostat
can fluctuate with the Boltzmann weight $\exp(-\beta {\cal H})$.
Therefore, the system will achieve the canonical ensemble when the system is ergodic.

\section{Simulation}

In order to investigate the ergodicity of the system with the Nos\'e-Hoover thermostat,
we study the following two systems,
\begin{eqnarray}
{\cal H}_{\mbox{ho}} &=& \frac{1}{2}(p^2 + q^2),\\
{\cal H}_{\mbox{ym}} &=& \frac{1}{2}(p_1^2 + p_2^2) + \frac{1}{2} q_1^2 q_2^2,
\end{eqnarray}
where ${\cal H}_{\mbox{ho}}$ and ${\cal H}_{\mbox{ym}}$ are the 
Hamiltonian of the harmonic oscillator and the Yang-Mills type, respectively.

\begin{figure}
\centerline{%
\includegraphics[width=.45\linewidth]{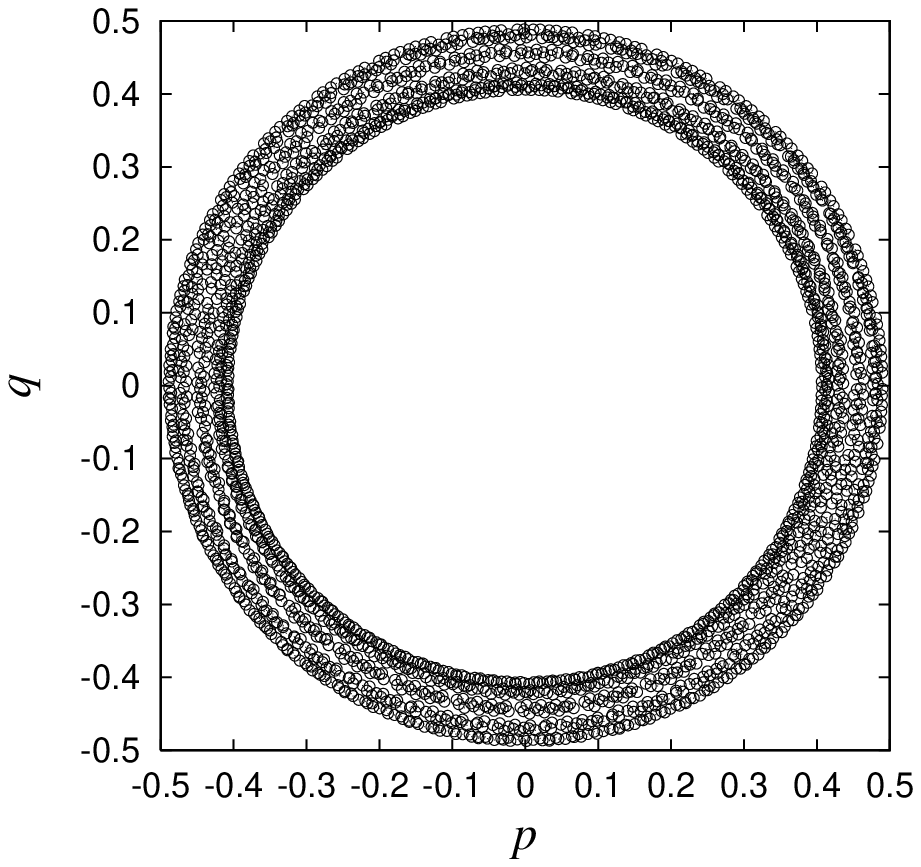}
\includegraphics[width=.45\linewidth]{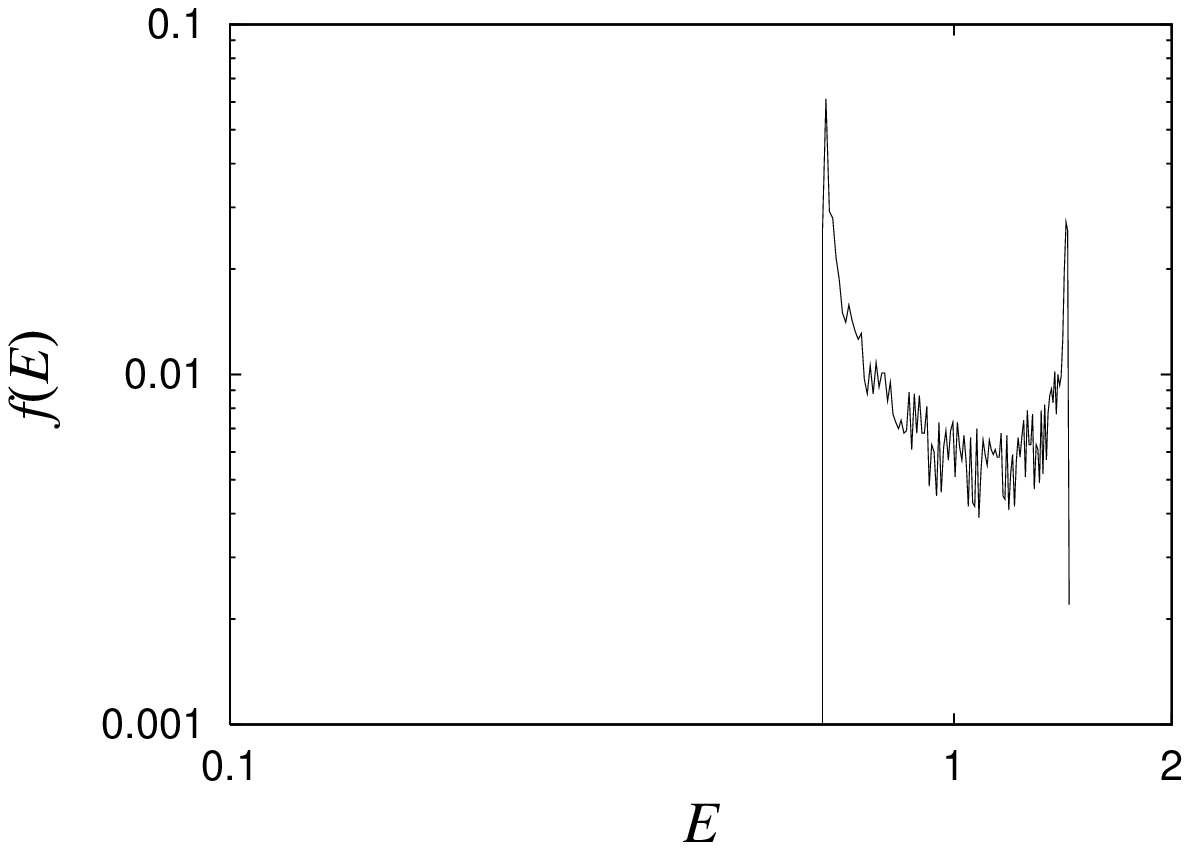}
}
\caption{(a) Phase diagram and (b) an energy histogram of the harmonic-oscillator system.
}
\label{fig_ho}
\end{figure}

We performed numerical simulations of the two systems with the Nos\'e-Hoover thermostat.
The time evolution of the systems were calculated by the fourth-order Runge-Kutta method.
The time step $\Delta t$ is set to be $0.005$, the aimed temperature $K_0$ is $0.5$, and the total step is $10^7$.
The results for the harmonic oscillator are shown in Fig.~\ref{fig_ho}.
This figure shows that the system loses the ergodicity, and the distribution 
of the energy is far from that of the canonical ensemble.

\begin{figure}
\centerline{%
\includegraphics[width=.45\linewidth]{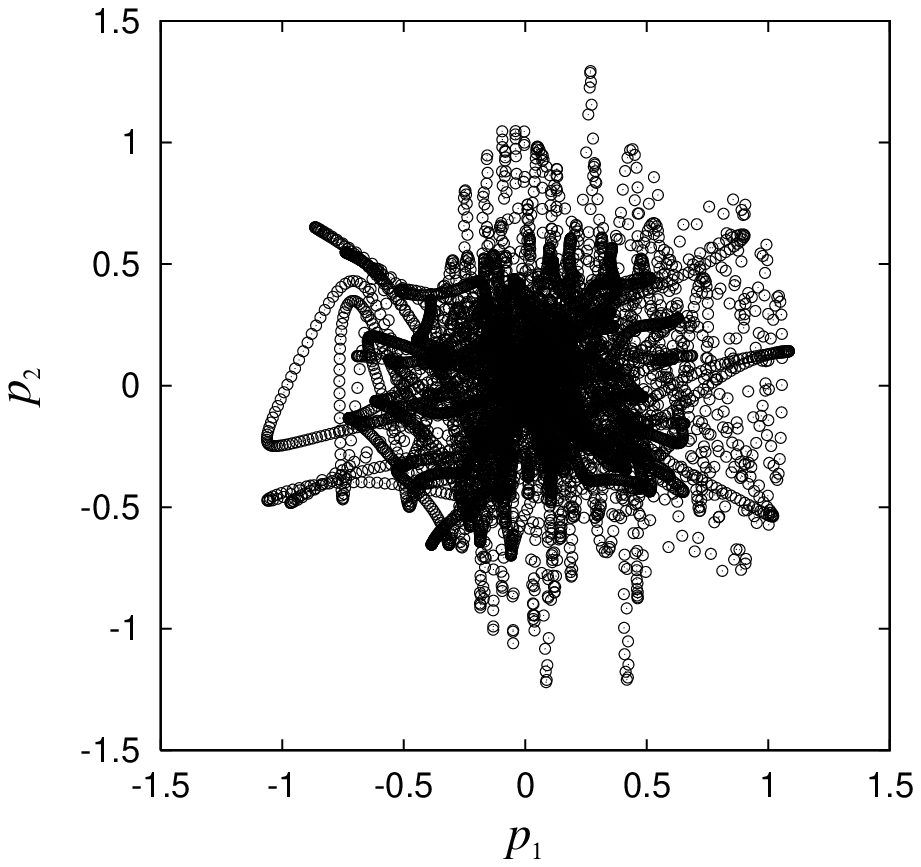}
\includegraphics[width=.45\linewidth]{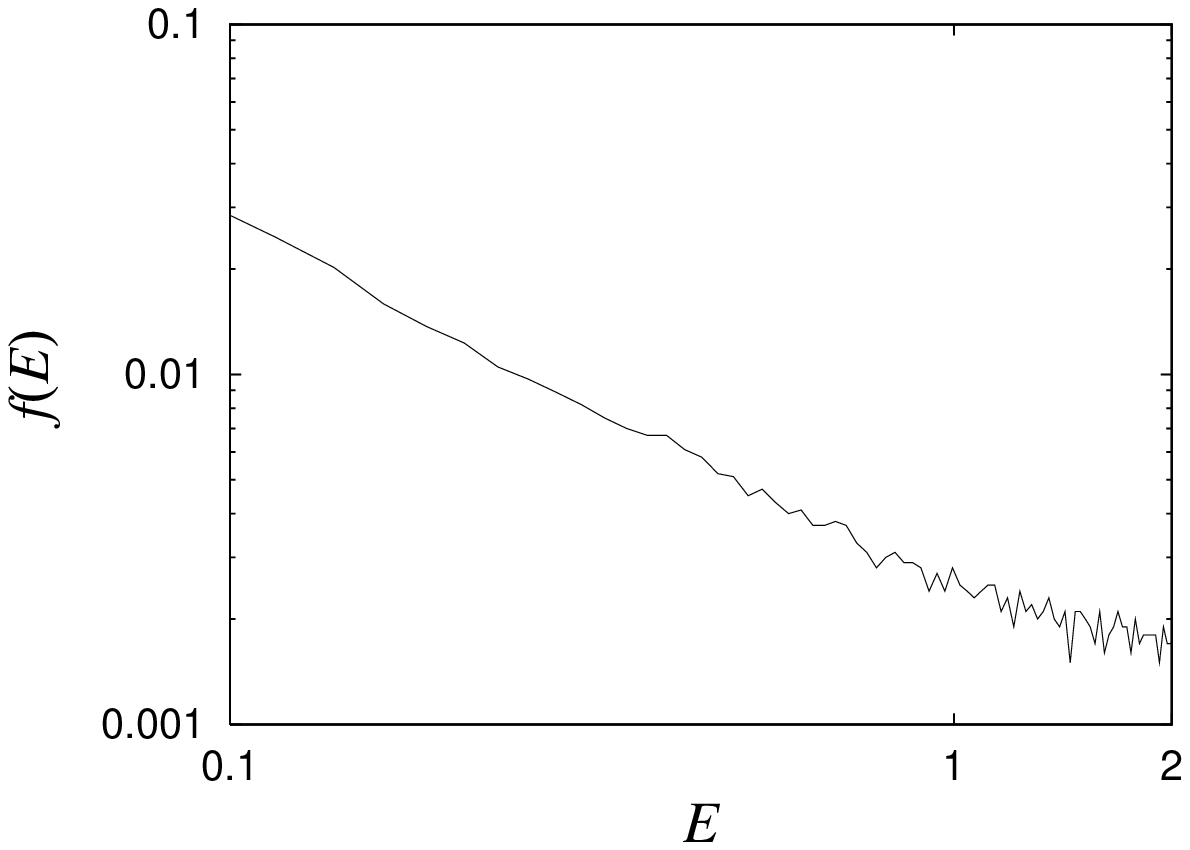}
}
\caption{(a) Phase diagram and (b) an energy histogram of the Yang-Mills system.
}
\label{fig_ym}
\end{figure}

The results of the Yang-Mills system are shown in Fig.~\ref{fig_ym}.
The phase diagram of $(p_1,p_2)$ is chaotic, and the histogram
of the energy shows canonical distribution, while the harmonic-oscillator system does not.
It suggests that the Yang-Mills system with the Nos\'e-Hoover thermostat
is ergodic while it has only two degrees of freedom.

%------------------------------------------------------------------------
% "M—'ÌŒ‹‰Ê
%------------------------------------------------------------------------

\begin{figure}
\centerline{%
\includegraphics[width=.45\linewidth]{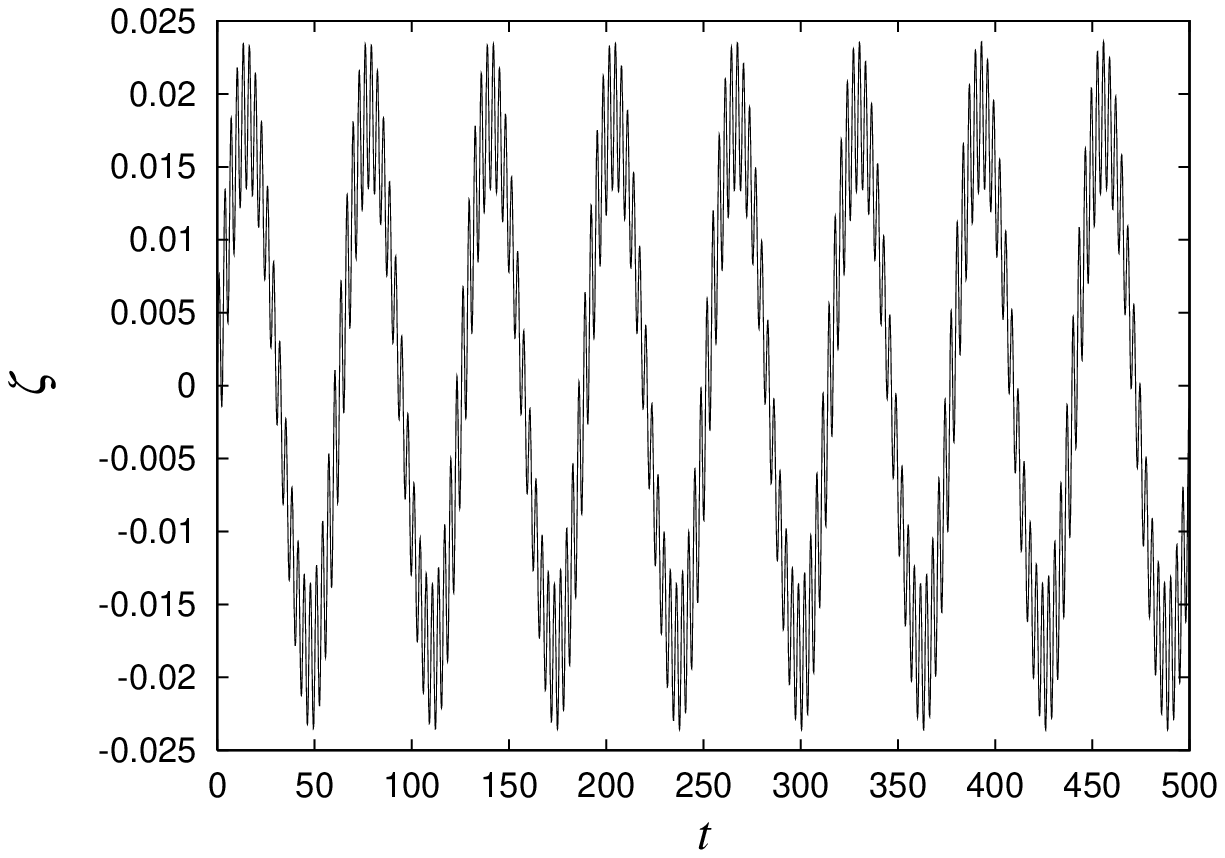}
\includegraphics[width=.45\linewidth]{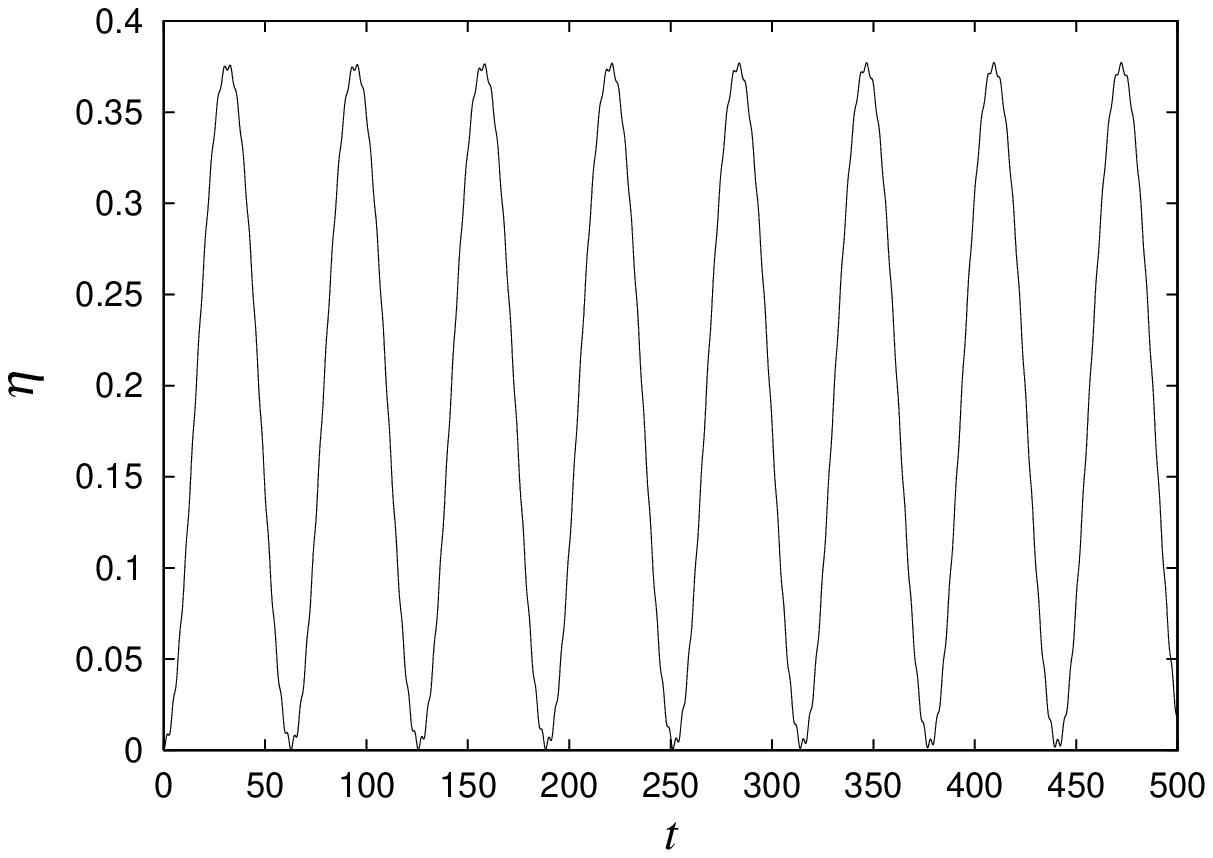}
}
\caption{%
The harmonic-oscillator system.
Time evolutions of the heatbath variables (a) $\zeta$ and (b) $\eta$ are shown.
}
\label{fig_ho_hb_variables}
\end{figure}

We focus the dynamics of the heatbath variables in order to figure out why the two systems show different behaviours.
The time evolution of the heatbath variables for the harmonic-oscillator system
is shown in Fig.~\ref{fig_ho_hb_variables}. The both variables seems to be
periodic, and the absolute values of them are bounded throughout the simulations.

\begin{figure}
\centerline{%
\includegraphics[width=.45\linewidth]{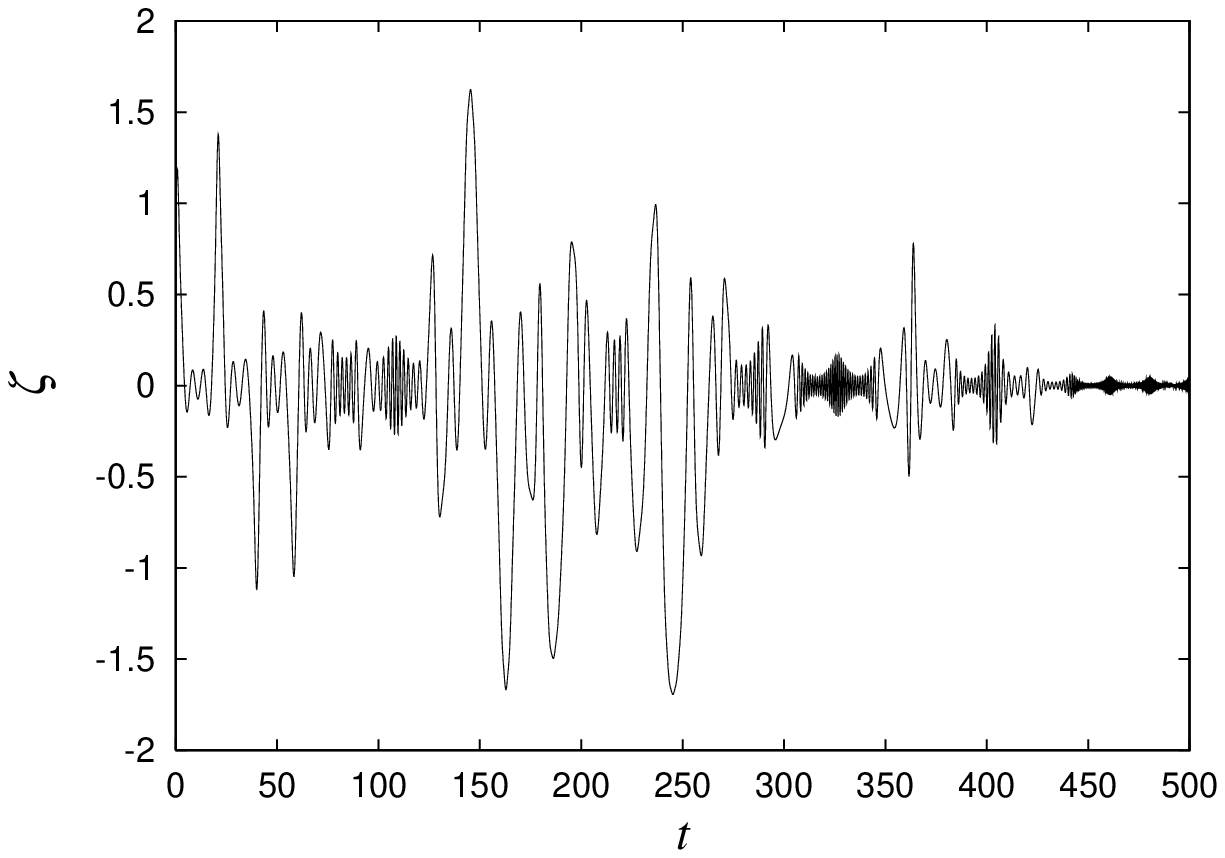}
\includegraphics[width=.45\linewidth]{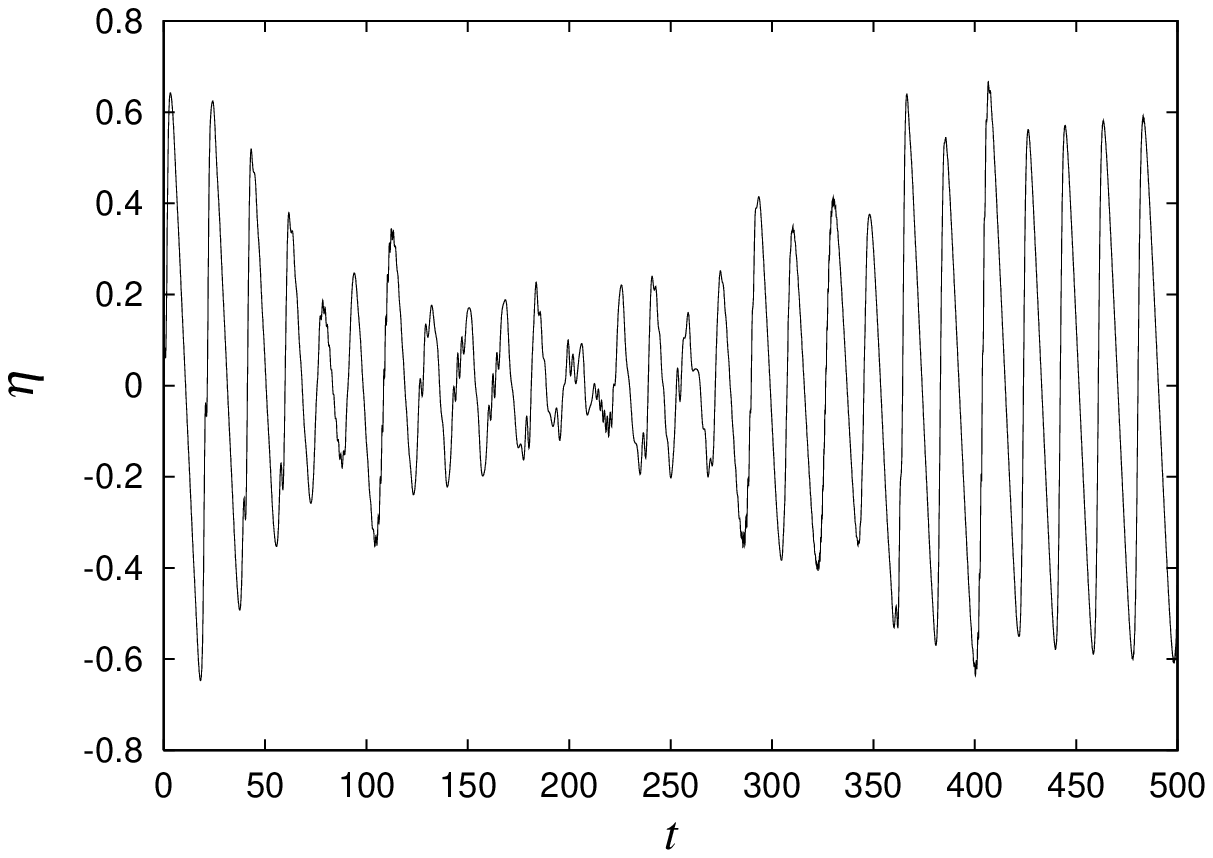}
}
\caption{%
The Yang-Mills system.
Time evolutions of the heatbath variables (a) $\zeta$ and (b) $\eta$ are shown.
}
\label{fig_ym_hb_variables}
\end{figure}

The time evolution for the Yang-Mills system is shown in Fig.~\ref{fig_ym_hb_variables}.
The behaviour of the heatbath variables seems to be chaotic, and
the amplitude is much larger compared to the harmonic-oscillator case.
The chaotic behaviour of the heatbath variables is essential for the
ergodicity of the system with Nos\'e-Hoover thermostat.

%------------------------------------------------------------------------
\section{Dynamics of Heatbath variables}
%------------------------------------------------------------------------

In this section, we derive the differential equation for the heatbath variables
in order to investigate why the variable show the periodic behaviour
in the harmonic-oscillator system.
The equations of motion Eqs.(\ref{eq_nh_motion_q}), (\ref{eq_nh_motion_p}),
and (\ref{eq_nh_motion_zeta}) are transformed to be
\begin{equation}
\ddot{\zeta} + 2 \dot{\zeta} \zeta + \frac{2}{\tau^2} \zeta = - \frac{1}{K_0\tau^2} pq,
\end{equation}
in the harmonic-oscillator case.

With large enough $\tau$, the heatbath variable $\zeta$ varies much slower than $pq$.
Therefore, we can take an approximation $\left< pq \right> = 0$.
Assuming the scaling form for $\zeta$ to be
\begin{equation}
\zeta(t) = \frac{1}{\tau} \bar{\zeta}(t/\tau), \label{eq_zeta_scaling}
\end{equation}
we have the differential equation for $\bar{\zeta}$ to be
\begin{equation}
\ddot{\bar{\zeta}} + 2 \dot{\bar{\zeta}} \bar{\zeta} + \bar{\zeta} = 0, \label{eq_zetabar}
\end{equation}
which is independent of $\tau$. Therefore, $\tau$-dependence of $\zeta$ should
have the form in Eq.~(\ref{eq_zeta_scaling}).
A quantity 
\begin{equation}
H_\zeta = \bar{\zeta}^2 + \dot{\bar{\zeta}} - \log(\dot{\bar{\zeta}}+1)
\end{equation}
is conserved in Eq.~(\ref{eq_zetabar}).
We can prove that $\bar{\zeta}$ and $\dot{\bar{\zeta}}$ are bounded 
because of $\dot{\zeta}+1 = K/K_0 > 0$.
Therefore, the variable $\bar{\zeta}$ should be periodic.
Note that $\eta$ should also be periodic since $\bar{\zeta}$ is an odd function.

\begin{figure}
\centerline{%
\includegraphics[width=.5\linewidth]{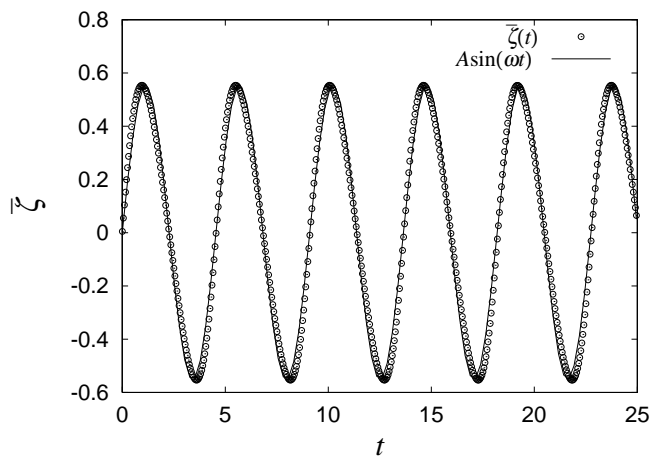}
}
\caption{
Time evolution of $\zeta$. 
The initial condition is $\bar{\zeta}|_{t=0} = 0, \dot{\bar{\zeta}}|_{t=0} = 1$.
The behaviour of the heatbath variable is periodic.
}
\label{fig_zeta}
\end{figure}

We solved Eq.~(\ref{eq_zetabar}) numerically using the fourth-order Runge-Kutta method.
The initial condition is $\bar{\zeta}|_{t=0} = 0, \dot{\bar{\zeta}}|_{t=0} = 1$.
The results are shown in Fig.~\ref{fig_zeta}.
The behaviour is periodic as expected.

\section{Kinetic-Moments Method}

Kinetic-moments method is one of the methods proposed to improve the ergodicity even for the harmonic-oscillator case \cite{KineticMoments}.
This method controls the kinetic energy $\left< K \right>$ and 
its fluctuation $\left< K^2 \right> - \left< K \right>^2$ simultaneously.
The equations of motion in this method are
\begin{eqnarray}
\dot{q} &=& \frac{\partial {\cal H}}{\partial p}, \\
\dot{p} &=& - \frac{\partial {\cal H}}{\partial q} - \zeta p - \xi p^3, \\
\dot{\zeta} &= & \left(p^2 -1 \right) /\tau_\zeta, \\
\dot{\xi} &= & \left(p^4 -3p^2 \right) /\tau_\xi,
\end{eqnarray}
with an additional heatbath variable $\xi$.
While Hoover and Holian reported the system with the kinetic-moments
thermostat shows good canonical distribution,
they did not explain why the method achieves ergodicity.

\begin{figure}
\centerline{%
\includegraphics[width=.45\linewidth]{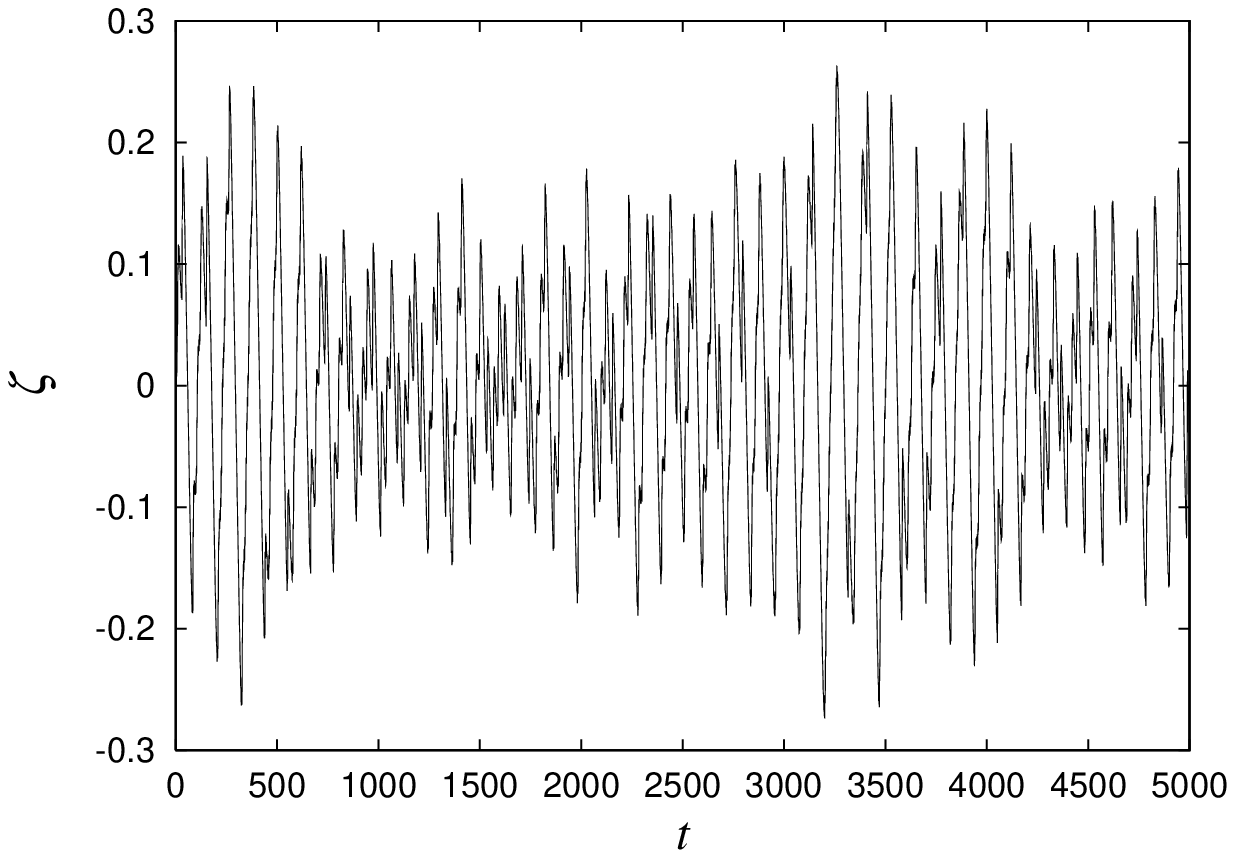}
\includegraphics[width=.45\linewidth]{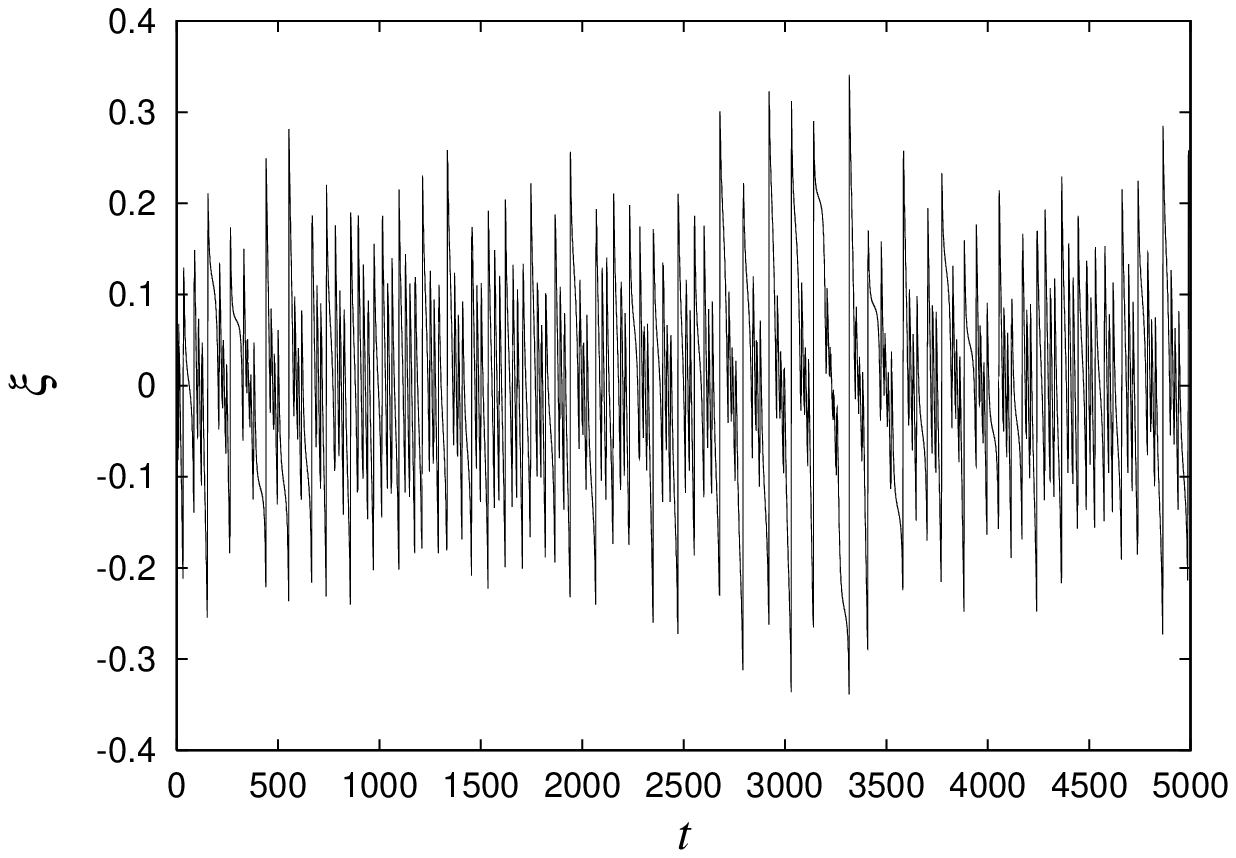}
}
\caption{Time evolutions of the heatbath variables.
(a) Time evolution of $\zeta$ and (b) that of $\xi$ are shown.
The behaviour seem to be chaotic and the amplitudes of the 
fluctuation are much larger than that of the Nos\'e-Hoover system.
}
\label{fig_km_hb_variables}
\end{figure}

Here, we discuss the behaviour of the heatbath variables in the method.
We simulated the harmonic-oscillator system with the kinetic-moments
thermostat. The details of the simulation are same as the previous.
The time evolution of the two heatbath variables $\zeta$ and $\xi$ are 
shown in Fig.~\ref{fig_km_hb_variables}.
The heatbath variables seem to be chaotic while the variables
are periodic in the Nos\'e-Hoover system.

We can derive the differential equations for the heatbath variables
in the kinetic-moments case. The equations are complex,
and are expected not to have any conserved values.
Therefore, the kinetic-moments method achieves the ergodicity
because of the chaotic behaviour of the heatbath variables,
not by controlling higher moments of the kinetic energy.

\section{Summary and Discussion}

We have studied the ergodicity of the Nos\'e-Hoover thermostat in 
the two systems, the one-dimensional harmonic-oscillator and Yang-Mills systems.
The numerical results have shown that the Yang-Mills system is ergodic
while the harmonic-oscillator system is not.

In order to study why the two systems have shown the different behaviour,
we have derived the differential equation for the heatbath variable.
When the system oscillates quickly, the heatbath variable becomes periodic,
and consequently the whole system cannot be ergodic.
Therefore the Nos\'e-Hoover system loses its ergodicity 
not because the system has a small degrees of freedom,
but the heatbath variables becomes periodic.
A particle system with the Nos\'e-Hoover thermostat 
will be ergodic, since quick oscillations are not expected to exist,
even if the system has small degrees of freedom.
We also explain $\tau$-dependence of the heatbath variable
in the amplitude and the period at the same time.

%---
% Kinetic Moments method
%---

We have also studied the ergodicity of the kinetic-moments method.
This method contains an additional heatbath variables to the Nos\'e-Hoover method,
and it makes the heatbath variables chaotic. The chaotic behaviour of the variables
achieves the ergodicity of the whole system.
These results imply that at least one of the original system and 
the heatbath variables should be chaotic in order that the whole system
is ergodic.

\section*{Acknowledgements}

We thank Miyashita, Ito, and Todo groups for fruitful discussion.
We also thank Sasa group for useful suggestions.
This work was partially supported by the 21st  COE program,
``Frontiers of Computational Science," Nagoya University.

\end{document}